\documentclass{collective-intelligence}

\usepackage{graphicx}

\begin{document}
\bibliographystyle{agsm}

\title{AN EXISTING, ECOLOGICALLY-SUCCESSFUL GENUS OF \\
COLLECTIVELY INTELLIGENT ARTIFICIAL CREATURES}

\numberofauthors{1} %  in this sample file, there are a *total*
\author{
% 1st. author
\alignauthor
Benjamin Kuipers \\
       \affaddr{University of Michigan}\\
       \affaddr{Computer Science \& Engineering}\\
       \affaddr{2260 Hayward Street}\\
       \affaddr{Ann Arbor, Michigan 48109}\\
       \email{kuipers@umich.edu}
}
\maketitle
\begin{abstract}
People sometimes worry about the Singularity \cite{Vinge-wer-93,Kurzweil-05}, or about the world being taken over by artificially intelligent robots.  I believe the risks of these are very small.  However, few people recognize that we {\em already} share our world with artificial creatures that participate as intelligent agents in our society:  corporations.  Our planet is inhabited by two distinct kinds of intelligent beings --- individual humans and corporate entities --- whose natures and interests are intimately linked.  To co-exist well, we need to find ways to define the rights and responsibilities of both individual humans and corporate entities, and to find ways to ensure that corporate entities behave as responsible members of society.
\end{abstract}

\section{Corporations are Intelligent Agents}

A corporation is an artificial legal entity, created by the state through a particular kind of legal agreement.  A corporation can own property, can sign contracts, can sue and be sued in court, and can be prosecuted and punished for crimes.  It can act as an economic {\em agent} on its own behalf in our society.

A corporation can have goals, can make plans to achieve those goals, and can use its resources to act to carry out those plans.  It solves problems and makes decisions about how best to achieve its goals, so it can be considered as an {\em intelligent agent}, as defined by a leading text in Artificial Intelligence \cite[p.~34]{Russell+Norvig-10}.

\begin{quote}\sf
An {\bf agent} is anything that can be viewed as perceiving its {\bf environment} through {\bf sensors} and acting upon that environment through {\bf actuators}.  . . .   A human agent has eyes, ears, and other organs for sensors and hands, legs, vocal tract, and so on for actuators.  A robotic agent might have cameras and infrared range finders for sensors and various motors for actuators.  A software agent receives keystrokes, file contents, and network packets as sensory inputs and acts on the environment by displaying on the screen, writing files, and sending network packets.   . . . 

We will see that the concept of rationality can be applied to a wide variety of agents operating in any imaginable environment.  Our plan in this book is to use this concept to develop a small set of design principles for building successful agents --- systems that can reasonably be called {\bf intelligent}. 
\end{quote}

\subsection{Knowledge and Inference}

In addition to having sensors and effectors, an intelligent agent must represent knowledge and perform inference in order to make rational decisions about its actions.

For a corporate entity, much of its knowledge is represented in external memory --- databases, documents, reports, and memos.  However, the majority of its knowledge is represented in the minds and in the skills of its constituent human members.  This includes knowledge of the ``corporate culture'' --- the way things are done in and by this particular corporation.  The corporate culture specifies how an individual must behave to function successfully as part of this particular corporate entity.

Inference within a corporate entity takes place, in part, through specified business processes.  Other kinds of inference take place, starting with an individual human decision, which is then evaluated by an appropriate set of members in the corporate entity against the goals, priorities, and values of the corporation, to determine the correctness of that conclusion or decision, before action is taken.

\subsection{The Genus of Corporate Entities}

A discussion of ``corporations'' typically focuses on for-profit corporations, since these have distinctive legal properties and are influential in our society.  However, we will treat {\em for-profit corporations} as one species within the larger genus\footnote{A biological taxonomist might argue that corporate entities should be considered to be a separate {\em kingdom}, rather than a mere {\em genus}, but we will leave this issue for future work.} of {\em corporate entities}, which also includes non-profit corporations, unions, governments large and small, churches old and new, and other species of corporate entities.

We define a {\em corporate entity} as a legal construct, constituted out of other intelligent agents, capable of acting as an agent on its own behalf.  That is, it can own property, can enter into contracts, can sue and be sued in court, and can be prosecuted and punished for crimes.

A corporate entity is made up of individual people (and of other agents), somewhat as an individual human is made up of cells, tissues, and organs.  The existence of a corporate entity is typically not limited to a specified period of time, or to the lifetime of any individual or set of human members, so it is potentially immortal.  

Corporations are not invulnerable, of course, since they can die of starvation, merge with or be devoured by others, or be destroyed in some other way.  Indeed, corporate entities exist and evolve in a world of Darwinian selection, which naturally leads to a survival drive.  (A robot, on the other hand, is a human-designed artifact and it may or may not have a survival drive.)

Defining the precise boundaries of the genus of {\em corporate entities} is difficult and subtle, and we shall not attempt to settle it here.  However, for many species of corporate entity (including for-profit corporations), the existence of an entity is straight-forwardly tied to the existence of a legal corporate charter.  

The degree of independent agency a corporation has, relative to particular individual human members, varies along a wide spectrum.  The agency of a tiny corporation is effectively indistinguishable from that of its human creators.  A large corporation such as Apple, Microsoft, or GM is a distinct agent from even its most powerful individual human members, much as a human being is a distinct entity from his or her lungs or heart or even brain.  This is clearly illustrated by the fact that Apple Computer, Inc.~was able (however unwisely) to fire its founder, Steve Jobs.  If they were the same agent, this would not have been possible.

The focus of this essay is on large, persistent corporate entities whose agency is clearly distinct from that of any given individual human.  While the boundaries may be controversial, the existence of the core cases is not.

\subsection{A corporation is not a person}

``{\em Corporations are people!\/}'', Mitt Romney, 8-11-2011.

A corporation is made up of people, somewhat as individual humans are made up of cells, tissues, and organs.  Few large corporations are under the control of any particular individual.  Even the founder or CEO of a large corporation keeps his position only as long as his or her actions are seen as consistent with the goals of the corporation.  

The 1985 decision by Apple Computer, Inc.~to fire Steve Jobs could be seen as the outcome of a political process among the very human senior managers at Apple.  However, this same process can also be seen as the mechanism by which a corporate entity makes a decision.  An analogy is with the neuronal balance between stimulatory and inhibitory neural activations that leads to winner-takes-all activation of a particular down-stream neuron \cite{Koch-hn-85}.

The checks and balances among the ``organs'' of a corporation do a pretty good job of preventing even a powerful individual like a founder, the CEO, or members of the Board of Directors from substantially changing the priorities of a corporation they are part of.  

Corporations are intelligent agents that function as part of our society.  However, it is important to be clear:
\begin{quote}\bf A corporation is not a person.  \\  
Individual human beings are persons. \end{quote}

The term ``legal person'' is a source of confusion.  
While it is a specialized term in law, the difference between the legal term and the commonsense understanding of ``person'' is so great that the use of the term is misleading.  A less misleading term for a corporation might be ``legal agent''.

The status of corporations as ``legal persons'' is usually attributed to a court decision \cite{Santa-Clara-County-1886}, but that assertion appears only in the ``headnote'' --- commentary by the court reporter without legal status \cite[Chap.~1]{Hartmann-10}.  Nonetheless, subsequent court decisions have built on this foundation, however inadequate.

\subsection{People and corporations are different}

An individual human person has consciousness and mental and emotional states, and adult humans can take responsibility for their actions.  To the extent that these concepts are applicable at all, they have very different meanings for corporations.

While a corporation can have goals, plans, and actions of its own, it seems unlikely that a corporation can be considered {\em conscious} in any meaningful sense \cite{Kuipers-aim-08}.  Without the ability to feel things like pain, or fear, or shame, or guilt, the concept of ``taking responsibility'' cannot mean for a corporate agent anything like what it means for an individual human being.  Therefore, a corporate entity, as such, does not have a {\em conscience}:  the ability to understand, feel, and regret what they have done wrong.

A corporate entity, as such, also lacks an {\em imagination}:  the ability to envision some state of affairs that does not yet exist.  For example, one of Steve Jobs' great gifts at Apple Computer, Inc.~was his ability to envision the qualities of the products he wanted designed, and to insist that product designs achieve his vision.  While there is AI research on innovative design, the capabilities of individual humans are far beyond what can be achieved by artificial computational systems or collectively intelligence corporate entities.  (``Design by committee'' is justly reviled.)

A human person is an individual, and is not the same person as any other individual.  (We will ignore extremely rare pathologies such as split-brain cases, conjoined twins, and multiple-personality disorders.)  Creation of a new person is an arduous (though not unpleasant) biological, educational, and social process.  The number and identity of individual human persons is therefore a reasonably stable property, allowing rights and responsibilities to be assigned in a reasonably stable way.

New corporations, however, can be created with relative ease, and they can split, merge, and partially or entirely own each other in ways that have no analog for individual humans.  ``Dummy'' corporations can easily be created to shield the identity and responsibility of individual or corporate actors.  This difference in properties is very important when considering the rights and responsibilities of the different types of agents in society.

\subsection{Non-humans own and control our world}

A large portion of the wealth and productive capacity of the world is controlled by corporations \cite{Hartmann-10}.  The effect of laws passed over the past 150 years or so have given corporations commanding advantages over individual human beings in the economic sphere.

The fundamental goal of a for-profit corporation is typically defined as increasing shareholder value\footnote{Note that ``increasing shareholder value'' really means increasing the assets, and therefore the power, of the corporate entity.  If the shareholders were to decide to ``cash out'' that value, for example by selling their shares in large numbers, the continued existence of the corporation itself is threatened.  An important survival goal for a corporation is to convince its shareholders to leave their share of the value in place.}, and progress toward this goal is interpreted and overseen by its Board of Directors.  (We will discuss this point in more detail below.)  Unlike individual human beings, a corporation does not have a bounded lifespan. This is important in the economic sphere, where the power of compound interest really takes off after several decades of economic life. 

Darwinian evolution ensures that surviving corporations are focused on accumulating and protecting wealth, which is life's blood to a corporation.  Not surprisingly, the economic sphere flows over into the political sphere. 

The U. S. Constitution gives the vote to individual humans, and not to corporations. However, corporations have accumulated large amounts of economic and political power, and they are driven to survive. Naturally, they respond to threats by taking actions to influence individual humans to vote in ways that protect their interests.

In the ``Citizens United'' decision \cite{Citizens-United-10}, the U. S. Supreme Court decided that corporations, as ``legal persons'', are entitled to free speech, just as the Constitution guarantees to individual human beings.  The explicit purpose of this is to allow corporations to spend their enormous economic resources to influence human voters without legal restrictions \cite[Chap.~11]{Hartmann-10}.

Large for-profit corporations like Exxon or Halliburton negotiate as peers with sovereign states, less powerful than a few, but more powerful than most.  In principle, they are governed by the laws of the country in which they are incorporated.  But in practice, they keep their headquarters in states or countries where governments and laws are congenial, and act globally.

\subsubsection{A dystopian view}

We live in a social, economic, and political ecology including two somewhat-interdependent species, each driven to survive, but each having different types of capabilities.  Humans do not face extinction, any more than do the chicken or the dairy cow, but rather, subjugation.

Consider the economic and political spheres of our world as areas of ongoing interaction and conflict between two different intelligent species that inhabit the Earth. Not conflict among races of human beings (who are equal under law and morality), but between genus {\em homo} and genus {\em corporation}, or at least the species of for-profit corporations.  Corporate entities are treated quite differently from humans under the law, and often to the disadvantage of humans \cite[Part~III]{Hartmann-10}.

Corporations are artificial entities, created by humans within constraints defined by human-made laws.  In principle, those laws could be changed.  However, corporations are now powerful members of the society, able to influence political decisions about changes to current law.  Do humans still have the power to change the laws, so that humans and corporations can co-exist in society, better than we do now?  

These concerns are real and serious.  Even so, I think the dystopian future is not inevitable.

\section{Corporations are members of our society}

In national political discussions, a wide range of opinions are expressed about whether corporations are good, job-creating entities to be supported and encouraged, or whether they are evil, exploitative entities that should be restrained or destroyed. 

People have a desire to work together.  People have a desire to identify with and work for something larger than themselves as individuals, whether it is family, church, nation, state, or corporation \cite{Maslow-68}.  The corporate form is one manifestation of this desire.  We may need to reconsider the specific legal constraints on that form, but eliminating corporations is not an option.

Furthermore, corporations are extremely wealthy and powerful.  It would be difficult in practical terms to exclude them from participation in our society.  

We must take seriously the idea that, both now and into the future, our society consists of two distinct kinds of beings --- individual humans and corporate entities --- whose natures and interests are intimately linked.  

As members of the same society as individual humans, corporations have responsibilities as well as rights.

\subsection{The Responsibility of a For-Profit Corporation}

It is often said that for-profit corporations are legally constrained to maximize shareholder value above all other considerations.  The claim is that boards of directors and senior executives can be successfully sued for pursuing goals other than immediate shareholder gain.  This is often supported by quoting the 1919 Dodge v.~Ford decision:
\begin{quote}\sf
A business corporation is organized and carried on primarily for the profit of the stockholders.  The powers of the directors are to be employed for that end.  The discretion of directors is to be exercised in the choice of means to attain that end, and does not extend to a change in the end itself, to the reduction of profits, or to the nondistribution of profits among stockholders in order to devote them to other purposes.  \cite{Dodge-v-Ford-19}
\end{quote}

Indeed, the founders of Craigslist were recently found delinquent in their fiduciary duty to a shareholder (eBay) for actions taken to protect their corporate culture at the cost of shareholder value.
\begin{quote}\sf
Having chosen a for-profit corporate form, the craigslist directors are bound by the fiduciary duties and standards that accompany that form. Those standards include acting to promote the value of the corporation for the benefit of its stockholders.  \cite[pp.~60--61]{eBay-v-Newmark-10}
\end{quote}

At the same time, several legal authorities \cite{Smith-jcl-98,Stout-ucla-07} argue that for-profit corporations may act in the interests of a number of different stakeholders, including employees, customers, creditors, the community, and the environment, as well as shareholders.  Such actions may be justified in terms of the {\em long-term} interests of the shareholders.

Court decisions on this topic tend to arise in the context of the defensive tactics taken by a board of directors to avoid a hostile takeover.  The question is whether a board may adopt tactics (e.g., a ``poison pill'') to repel a take-over, or whether it is obliged to seek and accept the highest offer.  The underlying concern is that members of a board of directors could act in their individual self-interest, rather than in the interests of the corporation or its shareholders.

The {\em business judgment rule} is the long-standing presumption that the actions of a board of directors are taken in the best interests of the corporation; that is, the burden of proof is on the contrary argument.  In the Unocal decision \cite{Unocal-v-Mesa-85}, the court narrowed this defense, requiring a board to show that the take-over constituted a threat to corporate policy or effectiveness, and that the response was reasonable in terms of impact on shareholders, creditors, customers, employees, and the community (the wider set of stakeholders is important here).  The next year, in the Revlon decision \cite{Revlon-v-MacAndrews-86}, the court held that, once the sale of a company is inevitable, the role of the other stakeholders falls away, and the board's responsibility is to get the highest price for the shareholders.

For the purposes of this essay, our main conclusion from this discussion is that it is {\em not illegal} for an on-going for-profit corporation to consider stakeholders other than the shareholders, and values other than maximizing profit, although these other considerations must be justifiable in terms of their long-term value for the shareholders.  For example, it would seem straight-forward for a large and long-lived corporation with substantial investments in agricultural land, or real estate in coastal cities, to justify significant investments toward solving the problem of climate change.

Of course, this point is not widely appreciated, and many corporations {\em do} make their decisions strictly to maximize profit and shareholder value.  This has led some commentators to assert that corporate agents are essentially psychopaths in the way they function in society \cite{Bakan-03}.  Our conclusion is that, while this may be true in some (and possibly many) cases, it is not {\em necessarily} true, and thus can potentially be changed.  It is both {\em possible} and {\em legal} for a for-profit corporation to act as a responsible member of society.

\subsection{A member of society has responsibilities}

Every person who participates in society has both rights and responsibilities.  Rights get a lot of attention, with much less given to responsibilities.

The law enforces certain responsibilities.  The obvious cases are prohibitions:  murder, theft, driving on the wrong side of the street, etc.  There are also laws against selfish inaction in serious situations:  reckless endangerment, negligent homicide, etc.

However, the responsibilities of a person in society go beyond simply obeying the law.  A person has ``moral responsibilities'' that are not specifically enforced by law, but which a person should do, in order to be a member in good standing of the society.  These are often expressed in ways like:
\begin{verse}\sf
    ``Ask not what your country can do for you -- ask what you can do for your country.''  \\
    ``Do unto others as you would have them do unto you.''  \\
    ``Love thy neighbor as thyself.''
\end{verse}
These involve acting in ways to support other members, and improve the society as a whole.

The Golden Rule --- ``{\sf Love thy neighbor as thyself\/}''  --- is taught in the Old Testament, in the New Testament, and in the writings of many other religions.  It means that it is not enough to obey the law, to avoid lying, stealing, killing, and so on.  We are instructed to  {\em love} our neighbors, which means to care about them and their well-being as persons, and not just treat them as instruments or obstacles in pursuit of our own goals.

If corporations are members of our society, then the Golden Rule means that corporations should treat individual humans as their neighbors, as persons whose own needs and interests matter, and not simply as instruments and obstacles to be optimized over, in the pursuit of wealth.

But if these moral responsibilities are not enforced by law, how can they be enforced?  What about the ``Mr.~Scrooge'' figure, who pursues his own interests with callous disregard, even contempt, for the welfare of others, but without breaking the law?  

{\em Social pressure} is the mechanism by which society reminds individuals that they are failing to live up to their moral responsibilities.  Other members of society communicate society's low regard for irresponsible behavior.  The individual comes to feel {\em shame} and {\em guilt}, which motivates them to change their behavior.

Lawrence Kohlberg \cite{Kohlberg-71} proposed a detailed stage theory of moral development, showing how individuals progress from simply avoiding punishment and pursuing self-interest, to responding to shame and guilt, to the development of a principled {\em conscience}.  

Many corporations are responsible and productive members of our society, creating wealth and jobs, taking care of their shareholders, customers, employees, the community, and the environment.  It is possible for corporations to be good neighbors as part of our society.

But many corporations are not good neighbors.  Some corporations act as though they have no responsibility to the society they are part of, that their only responsibility is to create wealth for their shareholders.  They seem to believe that society provides them with rights, but there are no corresponding responsibilities beyond avoiding legal prosecution.

This amounts to the ``pursuit of self-interest'' level near the bottom of Kohlberg's hierarchy of moral development.  Such a corporation attempts to avoid punishment for breaking the law, to the extent that punishment interferes with increasing shareholder value.  This can lead to corporate lobbying to change the law, or to treating legal penalties as simply part of the cost of doing business.

\subsection{Is the ``Invisible Hand'' Enough?}

Is it good for society to allow corporations' responsibilities to be only at the ``pursuit of self-interest'' stage of moral development?  The economic argument of the ``invisible hand'' \cite{Smith-1776}, says that it is.  The claim is that businesses, acting in their own self-interest, want more satisfied customers, and therefore compete to provide better value to those customers.

%A recent clear statement of this position:
%\begin{quote}
%	{\sf ``Each business is run for the benefit of its owners, its shareholders, its customers, and its employees.  It's not run for the benefit of the country.  That's not why people run businesses.''}  [Bill Frezza of the Competitive Enterprise Foundation, interviewed by Lynn Neary of NPR Morning Edition, 10-4-2011, 6:25 am.]
%\end{quote}

According to this position, it is perfectly reasonable and acceptable for a corporation to treat individual humans as instruments to be optimized and/or obstacles to be removed in order to maximize its wealth.  This behavior would be considered reprehensible, even immoral, for individual human beings.

In contrast, in his inaugural address, President John Kennedy said:  ``{\sf Ask not what your country can do for you -- ask what you can do for your country}.''  This view of citizenship reminds us that everyone who enjoys the benefits of a society, including the rights it confers, also bears corresponding responsibilities.  This principle should extend to corporations.

\subsubsection{Another dystopian future}

I have rejected as implausible the ``Terminator'' scenario, with hostile intelligent robots trying to exterminate the human race.  Nonetheless, AI and robotics are making steady progress, automating increasing numbers of jobs previously held by humans.

Historically, automation has created more jobs than it has destroyed.  However, those jobs are typically not for the same people, or for people with similar levels of training.  And there is no guarantee that the creation of more jobs will continue.  It is conceivable that all jobs could be eliminated, except for a few jobs for scientists, engineers, and senior managers \cite{Vonnegut-52}.

Taking this to the limit, one can imagine a corporation, perhaps in the financial industry, with no humans at all, workers or even shareholders!  If one takes a purely instrumental view of human work, this scenario becomes a plausible goal for corporations to work toward.

\section{What is responsible corporate behavior?}

What is responsible behavior for corporate entities?  Corporations in the modern sense have only existed for a few centuries \cite{Hartmann-10}, and we haven't had time to explore the answers to this question in the same depth as for individual humans.  The notion that corporations are only responsible for creating shareholder wealth is clearly inadequate, if they are to function as members of our society.

What is responsible behavior for individual humans?  We have been trying to answer this question for millennia.  There is no single clear answer, of course, but the Golden Rule --- ``{\sf Love your neighbor as yourself\/}'' --- appears in the writings of virtually all major religions.   More generally, treat other people the way you would want to be treated if you were in their circumstances.

\subsection{Responsibility is long-term self-interest}

The United States of America was created by its Founding Fathers based on a radical vision of the relationship between individual human citizens and their overall society \cite{Locke-1689}.  {\em Social contract theory} \cite{Hobbes-1651,Locke-1689} describes how government is justified by the agreement of the people to give up certain rights in return for overall improvements in welfare and security.

Today, we must reason carefully about the relationship between these AIs --- corporate entities --- and the society that we individual human beings share with them.

We need to find clear, coherent sets of rights and responsibilities for both corporations and individuals.  This is in the best interests of corporations, as well as of humans.  When we all act responsibly, we can vastly increase the prosperity and well-being of everyone in our society.  But if everyone, or even a significant minority, acts selfishly and irresponsibly, then much of the resources of society goes into attack and defense against each other, rather than into the growth of prosperity and well-being for the society as a whole.  We are all poorer as a result.

This illustrates an important aspect of responsibility, ethics, and even morality.  Very often, the long-term best outcome requires giving up short-term benefits, as in the famous Prisoner's Dilemma problem \cite{Axelrod-84,Poundstone-92}.

Historically, better pay for factory workers incrementally reduced profits, but created a much larger class of potential customers.  There was a cost in tax dollars for investments in infrastructure such as the Interstate Highway System, federally funded scientific research, and the Internet, but these unleashed growth and innovation that created vastly more wealth than the original investment.  As an almost whimsical example, agreeing that everyone should drive on the right side of the street modestly reduces everyone's freedom to drive where they wish, but vastly increases the efficiency and safety of driving for everyone.

It is sometimes said that socially responsible behavior is a waste of resources that will be driven to extinction by Darwinian competition among corporations \cite{Friedman-nytmag-70}.  However, rules for {\em corporate social responsibility}\footnote{\sf http://en.wikipedia.org/wiki/Corporate\_social\_responsibility}, like rules for ethical behavior by individuals, may encode non-local (and non-obvious) rules for optimal behavior in the long run.  Short-sighted corporations that {\em fail} to act responsibly may well find themselves selected against, just as states that have stuck to feudal forms of government have been marginalized.

\subsection{Irresponsibility leads to political unrest}

Both the Occupy Wall Street movement and the Tea Party movement are driven by a strong populist sense that the distribution of rights and responsibilities are seriously out of balance in our society.

Some people are offended that poor people might be getting a ``free ride'' through government supports, without taking responsibility for their own welfare.  Other people are offended that wealthy and powerful corporations might be getting a ``free ride'', getting bailed out at great cost after their irresponsible behavior brought the entire economic system to the brink of failure, and then acting only in their own self-interest, rather than helping the recovery of the society as a whole. 

There is likely truth behind both movements, and the balance between rights and responsibilities must be restored in both situations.  It is important, even essential, to be clear about the rights and responsibilities of everyone in our society.  We focus here on the rights and responsibilities of corporations because they control vastly greater wealth and power in our society than do poor people.

\subsection{Productivity versus jobs}

Maximizing {\em productivity} is a key goal for corporate economic agents, and is widely reported and applauded.  Since productivity is the ratio of the value of goods and services produced to the value of materials and work used to produce them, the wages paid to human workers appears in the denominator.   Therefore, productivity is often maximized by finding ways to decrease human jobs and human wages.

Jobs are important for individuals, in part so they can earn money to take care of their needs and wants, and so they can pay taxes to help support the general society.  However, it is just as important that, when a person holds a job, he or she is taking responsibility for doing that job, doing it well, and earning the money they receive.  The worker is learning how to take responsibility, and learning to appreciate the value of responsible behavior.  Without jobs, people become poor and dependent, sometimes criminal, a burden to society rather than contributors.  Lack of jobs contributes to the imbalance between rights and responsibility in individual humans.

Providing jobs is not simply a cost of production, to be minimized to increase productivity and profits.  Providing jobs is one way a corporation contributes to society as a whole \cite{Schumacher-79}.  Responsible corporations recognize this.  It is said that, at the beginning of the 20th century, Henry Ford paid his auto workers several times more than the prevailing wage so they could afford to buy his cars.  Far from cutting into corporate profits, this seemingly irrational act contributed to the creation of the American middle class and helped make both the United States and the Ford Motor Company wealthy and successful.

\subsection{Can responsible behavior be enforced?}

For society to function, there must be ways beyond enforcement of legal constraints to ensure responsible behavior.  Most people feel empathy for other people, which encourages behavior following the Golden Rule.  Indeed, it is argued that lack of empathy may be responsible for psychopathy and evil in people \cite{Baron-Cohen-11}, and others have extended the argument to corporations \cite{Bakan-03}.

Parents work to instill a sense of right and wrong in their children, often by teaching the appropriate occasions for feelings of shame and guilt \cite{Kohlberg-71}.  These feelings are important aversive responses to social pressure, and are thus important to the social control of bad individual behavior.

Without consciousness, it is hard to imagine how a corporation can have feelings like empathy, shame, or guilt.  (But see \cite{Hall-07}.)  We have discussed how a corporate entity can explicitly rely on the imagination and creativity of individual humans who are part of the corporation.  Perhaps there is a way to ensure that a corporate entity cultivates and uses the sense of moral responsibility in its human elements, rather than subordinating those moral responses to an overly simple corporate goal, such as ``Profit above all!''

\section{Conclusions}

Insights from many different fields will be needed to solve this complex problem.

{\em Artificial Intelligence} contributes the insight that artificial entities can be intelligent agents, and in particular, that corporate entities meet the criteria for being intelligent agents that participate in our society.  

{\em Ecology} contributes insights into the diverse ways that different species can co-exist in a shared ecosystem, cooperating and competing simultaneously.  {\em Evolutionary biology} contributes insights into the ways that such ecological relationships evolve over time, and how they influence, and are influenced by, the resources available in the environment.  {\em Sociology} and {\em anthropology} contribute insights into the diverse ways that societies of intelligent human agents interact with each other, including the meaning of {\em culture}.  In our perspective here, corporate culture is one of the ways that a corporate entity represents knowledge and maintains its identity.

{\em Business} and {\em law} contribute insights into the structure, behavior, and function of corporate entities.  {\em Economics} contributes insights into the primary environment within which corporate entities, particularly for-profit corporations, perceive, plan, and act.  An important insight from law is that a common observation --- that for-profit corporations must by their nature put profit above all other considerations --- is not as immutable as commonly believed.

The study of {\em moral philosophy} and {\em moral evolution} contributes insights into how rules for responsible, ethical, and moral behavior constrain action in ways that result in substantially better outcomes for the society as a whole, and in many cases for every individual within it.  The study of {\em moral development} provides insights into how these rules are learned by children and other newcomers into a society.

It is plausible that eventually, evolutionary processes will select for corporate entities that follow (and compel each other to follow) rules for responsible behavior that improve the state of everyone in society.  However, learning through evolution is a process that is slow, painful, and expensive in lives and resources.  Evolution operates through both competition and catastrophic environmental collapse, killing off vast numbers of individuals and species, depending on statistical improvements in fitness among the population of survivors.    

As intelligent creatures, both in genus {\em homo} and in genus {\em corporation}, we can hope to draw on the knowledge we already possess, and learn better ways to act and interact in our society.  Perhaps it will be possible to improve on the slow and unimaginably painful evolutionary process, and find ways for humans and corporate entities to co-exist in a positive way.

%\begin{figure*}

%\begin{center}
%\hrule
%\vspace{1mm}
%\includegraphics[width=4in]{corporations2.pdf}
%\end{center}

%\caption{Corporations}
%\vspace{1mm}
%\hrule
%\end{figure*}

%ACKNOWLEDGMENTS are optional
\section{Acknowledgments}

This work has taken place in the Intelligent Robotics Lab in
the Computer Science and Engineering Division of the University of Michigan.
Research of the Intelligent Robotics lab is supported in part
by grants from the National Science Foundation (CPS-0931474 and IIS-1111494), 
and from the TEMA-Toyota Technical Center.

\bibliography{corp-AI}  

\end{document}